\documentclass[10pt,aps,prd,reprint,superscriptaddress,
citeautoscript,
amsmath,amssymb,longbibliography
]{revtex4-2}
\usepackage[hidelinks,colorlinks,allcolors=blue]{hyperref} 
\usepackage[]{graphicx}
\usepackage{amsmath}
\usepackage[
mode=text,  
range-phrase=\textendash,
range-units=single,
retain-explicit-plus,
]{siunitx}
\usepackage[table]{xcolor}
\usepackage{booktabs}  
\usepackage[caption=false]{subfig} 
\usepackage{makecell,multirow}  
\usepackage{rotating,bm}   
\usepackage{mathptmx}  
\usepackage{enumitem}
\setlist{nosep} 
\usepackage[right,pagewise,modulo]{lineno} 
\newcommand*{\citen}[1]{\begingroup\romannumeral-`\x\setcitestyle{numbers}\cite{#1}\endgroup}
\usepackage{verbatim}

\usepackage[version=4]{mhchem} 
\usepackage{xspace}  
\usepackage{anyfontsize} 

\graphicspath{{figures-tables/}}  

\setlist[description]{leftmargin=0pt,labelindent=0pt}

\newcommand{\rtwoscan}{r\textsuperscript{2}SCAN\xspace}
\newcommand{\dfcx}{vdW-DF-cx\xspace}

\newcolumntype{P}[1]{>{\centering\arraybackslash}p{#1}}
\newcolumntype{M}[1]{>{\centering\arraybackslash}m{#1}}

\makeatletter
\newcommand*{\addFileDependency}[1]{
  \typeout{(#1)}
  \@addtofilelist{#1}
  \IfFileExists{#1}{}{\typeout{No file #1.}}
}
\makeatother

\usepackage[
a4paper, 
textwidth=18cm, 
top=1.8cm,
bottom=2.5cm,
]{geometry}

\usepackage{makecell}
\DeclareSIUnit\angstrom{\protect \text {Å}}

\usepackage[all]{hypcap}

\newcommand{\ourtitle}{An experimentally validated end-to-end framework for \textit{operando} modeling of intrinsically complex metallosilicates}

\setcitestyle{super}  

\usepackage{newfloat}
\DeclareFloatingEnvironment[
    fileext=lof,
    name={Extended Data Fig.},
    listname={List of Extended Data Figures}
]{extfigure}
\DeclareFloatingEnvironment[
    fileext=lot,
    name={Extended Data Table},
    listname={List of Extended Data Tables}
]{exttable}

\usepackage{pdfpages} 
\usepackage{pgffor} 

\makeatletter
\AtBeginDocument{\let\LS@rot\@undefined}
\makeatother

\def\supplementfilename{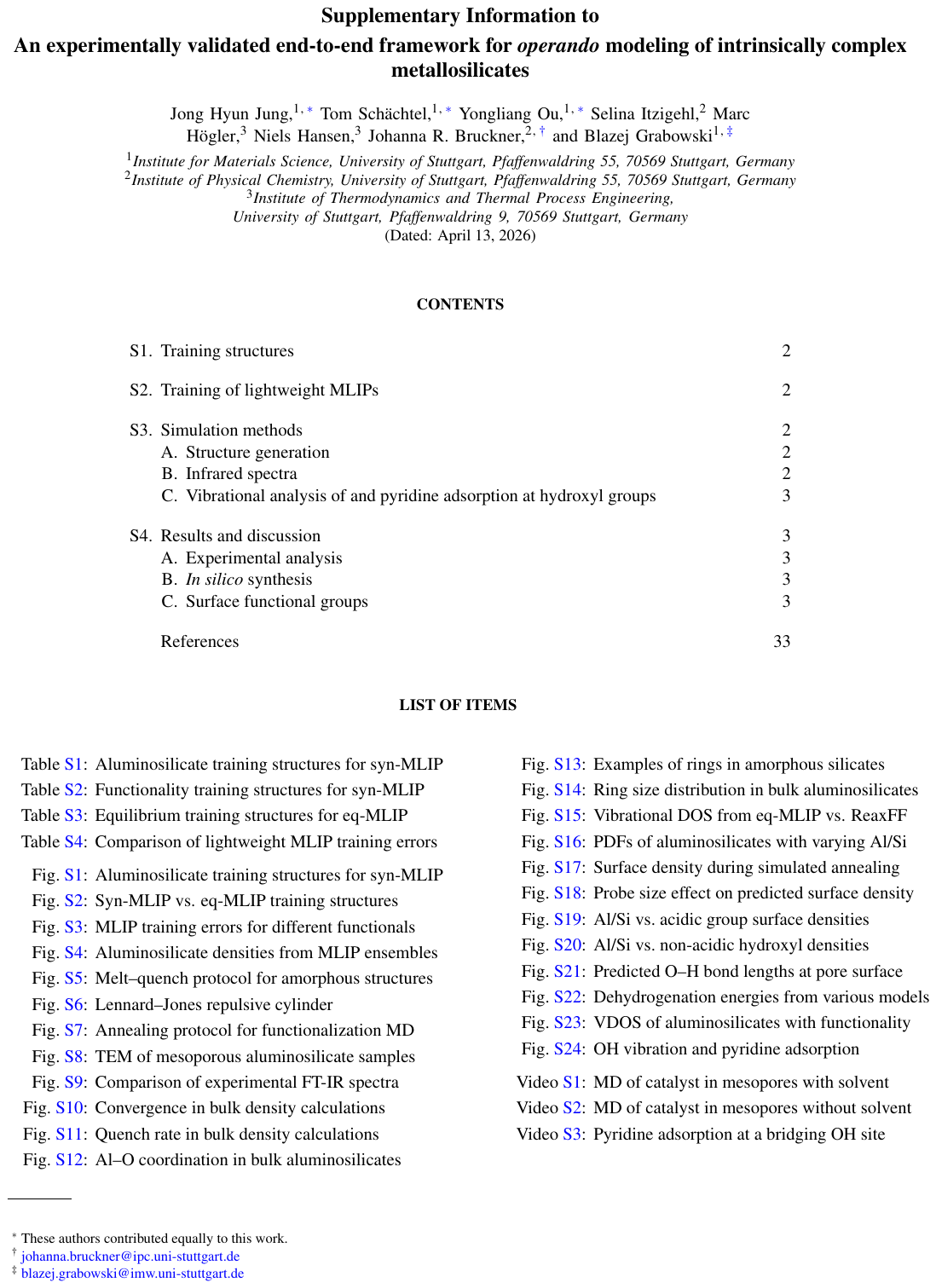}

\newif\ifarXiv
\arXivtrue

\begin{document}


\title{\ourtitle}

\author{Jong Hyun Jung}
\thanks{These authors contributed equally to this work.}
\author{Tom Sch\"{a}chtel}
\thanks{These authors contributed equally to this work.}
\author{Yongliang Ou}
\thanks{These authors contributed equally to this work.}
\affiliation{Institute for Materials Science, University of Stuttgart, Pfaffenwaldring 55, 70569 Stuttgart, Germany}
\author{Selina Itzigehl}
\affiliation{Institute of Physical Chemistry, University of Stuttgart, Pfaffenwaldring 55, 70569 Stuttgart, Germany}
\author{Marc Högler}
\author{Niels Hansen}
\affiliation{Institute of Thermodynamics and Thermal Process Engineering, University of Stuttgart, Pfaffenwaldring 9, 70569 Stuttgart, Germany}
\author{Johanna R. Bruckner}
\email{johanna.bruckner@ipc.uni-stuttgart.de}
\affiliation{Institute of Physical Chemistry, University of Stuttgart, Pfaffenwaldring 55, 70569 Stuttgart, Germany}
\author{Blazej Grabowski}
\email{blazej.grabowski@imw.uni-stuttgart.de}
\affiliation{Institute for Materials Science, University of Stuttgart, Pfaffenwaldring 55, 70569 Stuttgart, Germany}

\date{\today}

\begin{abstract}

Structurally and chemically complex materials such as amorphous metallosilicates underpin major catalytic and separation technologies, yet their intrinsic complexity challenges reliable atomistic modeling under realistic conditions.
Consequently, simulations that connect composition to material properties remain largely inaccessible for these materials.
Here, we enable quantitative \textit{operando} atomistic modeling of intrinsically complex materials through an experimentally validated end-to-end computational framework.
The approach combines separation of simulation domains, lightweight machine-learning potentials trained on high-fidelity data, and large-scale \textit{de novo} \textit{in silico} synthesis that mimics experimental procedures.
We apply the framework to realistic mesoporous \ce{SiO2(Al2O3)$_{x/2}$} ($0 \leq x \leq 0.4$) and validate the results experimentally.
Simulations quantitatively reproduce multiple experimental observables, including bulk densities, pair distribution functions, infrared spectra, and hydroxyl densities. 
Beyond prediction, the framework enables analysis of acid sites and vibrations for catalytic and adsorption processes.
By integrating simulation and experiment within a unified workflow, we advance the realism and reliability of atomistic modeling for intrinsically complex materials.

\end{abstract}

\maketitle

\section*{Introduction}

Metallosilicates represent a versatile class of intrinsically complex materials used in industry for catalysis~\cite{Fahda_Fayek_Dib_Cruchade_Pichot_Chaouati_Pinard_Petkov_Vayssilov_Mayoral_et_al._2024}, separation~\cite{Liu_Ge_Du_Song_Zhang_Zhou_Zhang_Gu_2024}, sensing~\cite{Sun_Romolini_Dieu_Grandjean_Keshavarz_Fron_D’Acapito_Roeffaers_Van_der_Auweraer_Hofkens_2024}, and energy technologies~\cite{Liu_Liu_Yang_Wu_2024}. 
These materials feature a structural hierarchy---from amorphous network topology and micro--meso--porosity to particle-scale grains---that enables tunable adsorption properties~\cite{Sheng_Zeng_2015, H.Hakim_H.Shanks_2011}.
In parallel, their high chemical flexibility, including incorporation of various transition metals and adjustable metal loading, permits control over interfacial acidity and catalytic activity~\cite{gajardo23,liang17}.
Yet, systematic experimental investigations across this vast structural--chemical design space remain costly and time-consuming. 

Atomic-scale modeling provides a powerful mechanistic route to uncover structure--property relationships and accelerate the rational design of metallosilicates. 
Reliable atomistic simulations, however, require access to highly accurate potential-energy surfaces.
Electronic-structure methods within the density functional theory (DFT) framework deliver such accuracy but are computationally prohibitive for large models and for extracting kinetic information through molecular dynamics (MD) simulations~\cite{Van_Speybroeck_Bocus_Cnudde_Vanduyfhuys_2023,tielens20}.
Empirical force fields such as ReaxFF~\cite{zhang24,rimsza16,pitman12}, calibrated against experimental or \textit{ab initio} data, enable large-scale MD but often are not adequate to describe multicomponent systems due to their fixed functional forms~\cite{erlebach22, erlebach2024}.
Recently developed high-dimensional foundation models~\cite{Batatia_Batzner_Kovács_Musaelian_Simm_Drautz_Ortner_Kozinsky_Csányi_2025, yang2024mattersim, deng23, Chen_Ong_2022}, including GRACE~\cite{PhysRevX.14.021036}, show promising transferability across broad chemical spaces. 
Still, their accuracy remains constrained by the quality and diversity of the available training data, which predominantly consist of near-equilibrium configurations~\cite{barroso_omat24,deng23}.
As a result, uncertainties increase for reactive events involving transition states or substantial bond rearrangements. These uncertainties are further compounded by the choice of the underlying DFT convergence parameters and exchange--correlation functionals. Commonly employed functionals (e.g., Perdew--Burke--Ernzerhof, PBE~\cite{barroso_omat24,deng23}) do not capture dispersion interactions adequately~\cite{forslund25,liu18,Nasir_Guan_Jee_Woodley_Sokol_Catlow_Elena_2025}. 
Fine-tuning and uncertainty quantification are also non-trivial owing to the high dimensionality of the parameter space.\cite{hanseroth2025}
Lightweight machine-learning interatomic potentials (MLIPs)~\cite{Yuan_Liu_Chen_Zhong_Raja_Kreiman_Vargas_Xu_Head-Gordon_Yang_etal._2026}, tailored to the targeted structural and chemical domains and enriched by active learning, offer a practical alternative: 
They require comparatively small training sets due to moderate model complexity, enable high-fidelity reference calculations, and remain accurate and computationally efficient for large-scale MD simulations. 

\begin{figure*}[]
   \centering
   \includegraphics{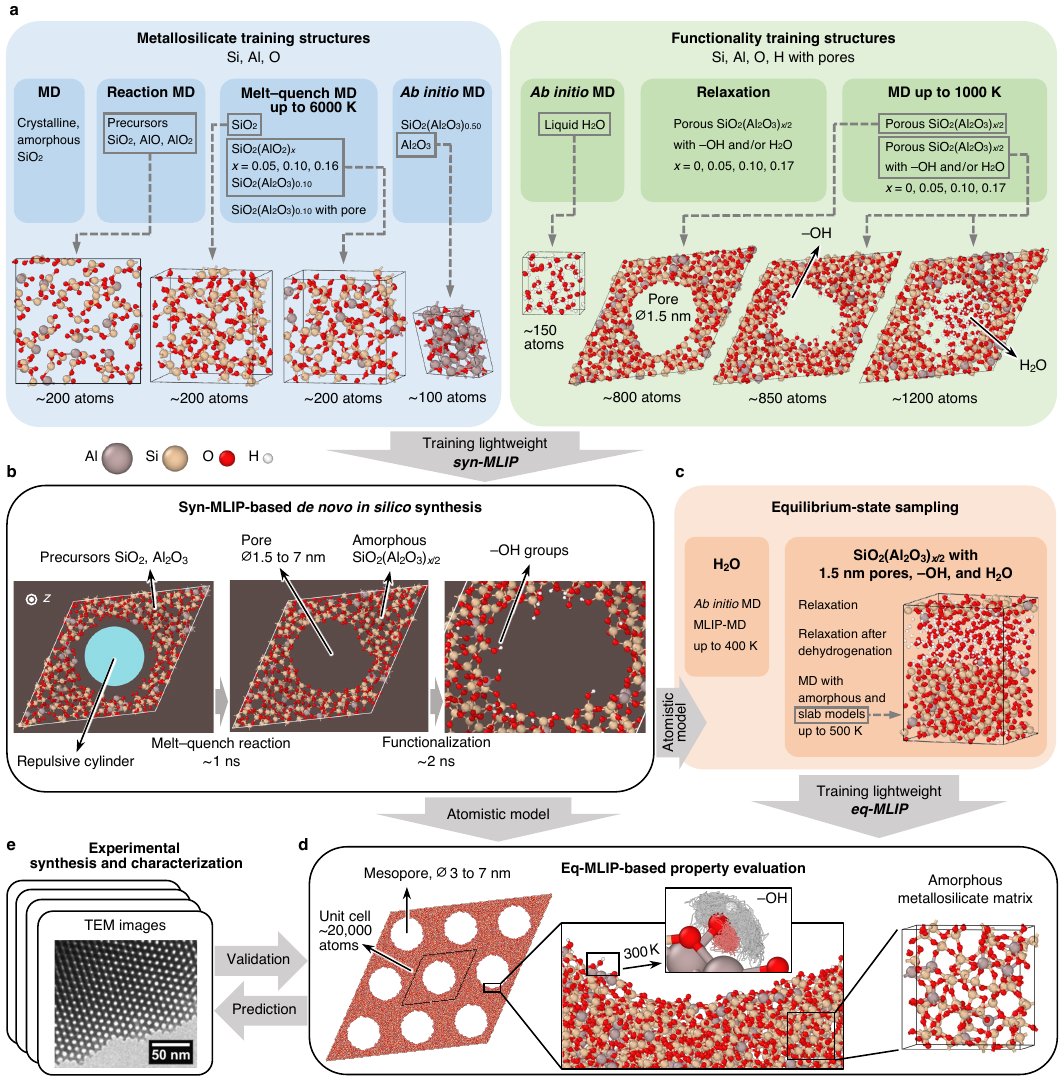}
   \caption{
   \textbf{Overview of the end-to-end framework.}
   \textbf{a},~Generation of lightweight syn-MLIP (machine-learning interatomic potential) training structures relevant to metallosilicates and functionality.
   Amorphous metallosilicates are synthesized \textit{in silico} via precursor reactions and melt--quench procedures in molecular dynamics (MD) simulations, starting from silica--metal oxide precursor mixtures.
   Active learning is performed in MD up to \qty{6000}{K} with syn-MLIP to explore the complex configurational space. 
   Additional structures are obtained from \textit{ab initio} MD. 
   The dataset is enriched with liquid water and H$_2$O--metallosilicate interface structures for functionality simulations. 
   \textbf{b},~Resolving the realistic experimental metallosilicate structures using syn-MLIP. 
   An artificial repulsive potential is applied to define a one-dimensional pore, followed by precursor filling and melt--quench amorphization.
   After removal of the repulsive potential, the pore surface is functionalized with hydroxyl (--OH) groups using syn-MLIP.
   \textbf{c},~Generation of eq-MLIP training set via equilibrium-state sampling.
   MD-driven active learning up to \qty{500}{\kelvin} is used to sample water-infiltrated metallosilicates, as well as slab and dehydrogenated structures.
   \textbf{d},~Large-scale MD simulations of realistic metallosilicates with functionality. 
   The lightweight and accurate eq-MLIP enables efficient evaluation of the target material properties. 
   As an illustrative example, we select a surface hydroxyl group and visualize its trajectory in MD at \qty{300}{\kelvin}.
   \textbf{e},~Experimental synthesis and characterization of metallosilicates, providing quantitative validation of the end-to-end framework. 
   Upon validation, eq-MLIP can be applied to predict a wide range of properties of complex metallosilicates. 
   Additionally, \textit{operando} modeling provides mechanistic insight to interpret experimental data.
   A representative transmission electron microscopy (TEM) image of the synthesized mesoporous silica is shown. 
   }
   \label{fig:workflow}
\end{figure*}

The key to developing lightweight MLIPs lies in constructing suitable training datasets.
The dataset must be rigorously curated to remain both representative and of manageable size, as excessive growth in size may ultimately compromise accuracy and training efficiency.
For metallosilicates, most available MLIPs targeted relatively simple structures (see previous studies in Extended Data Table~\ref{table:literature}), although complex motifs---such as mesoporosity and surface functionality---are essential for realistic applications~\cite{yu22,wei22}.
Developing MLIPs capable of describing such motifs is hindered by several factors: 
(i) mesopore-containing atomic structures are generally unknown \textit{a priori} and must be resolved \textit{in silico}; 
(ii) reactive chemistries involve mechanisms that remain largely unexplored; and
(iii) the chosen exchange--correlation functional must be validated, given the importance of dispersion interactions in metallosilicates.\cite{fischer19}
Furthermore, schemes integrating MLIP development and validation against experimentally measurable properties---rather than relying solely on regression metrics---are still lacking for metallosilicates, restricting the predictive reliability of prior simulation results (Extended Data Table~\ref{table:literature}).

Here, we propose an end-to-end computational framework based on lightweight MLIPs tailored to model structurally and chemically complex materials such as amorphous metallosilicates. 
Training datasets are generated via iterative active learning embedded in an end-to-end pipeline that spans \textit{de novo} \textit{in silico} synthesis, functionalization, and property evaluation.
A domain-specific training-and-deployment strategy is introduced in which multiple lightweight MLIPs are optimized to capture distinct domains of the potential-energy surface.
%

We apply the framework to prototypical ordered mesoporous metallosilicates \ce{SiO2(Al2O3)$_{x/2}$} ($0 \leq x \leq 0.4$), exhibiting enhanced hydrothermal stability and strong Brønsted acidity~\cite{gajardo23}.
The resulting datasets span a wide configurational space, covering variations in Al/Si composition, surfaces, pores, functional groups, water, and interfaces. 
Dispersion-aware functionals (\dfcx and \rtwoscan-D4) are used for labeling, due to the critical role of dispersion interactions in describing metallosilicates~\cite{Morawietz_Singraber_Dellago_Behler_2016,fischer19}. 
We also perform experimental validation for the end-to-end framework, quantitatively comparing key materials properties, including bulk densities, pair distribution functions (PDFs), infrared spectra, and functional group densities. 
Finally, we demonstrate that the framework enables acid-site and vibration analysis, thereby accelerating rational design of metallosilicates for catalysis and adsorption applications.

\section*{Results}

\subsection*{End-to-end learning and modeling}

Figure~\ref{fig:workflow} presents an overview of the end-to-end computational framework, which starts from precursors, resolves the bulk and surface structures of porous metallosilicates, and predicts experimentally measurable properties with near-\textit{ab initio} accuracy, by combining lightweight MLIPs with large-scale MD simulations.
A domain-specific MLIP training strategy is integrated to complement the end-to-end framework, delivering accuracy and efficiency across all stages of modeling.

\begin{figure*}[]  
   \centering  
    \includegraphics{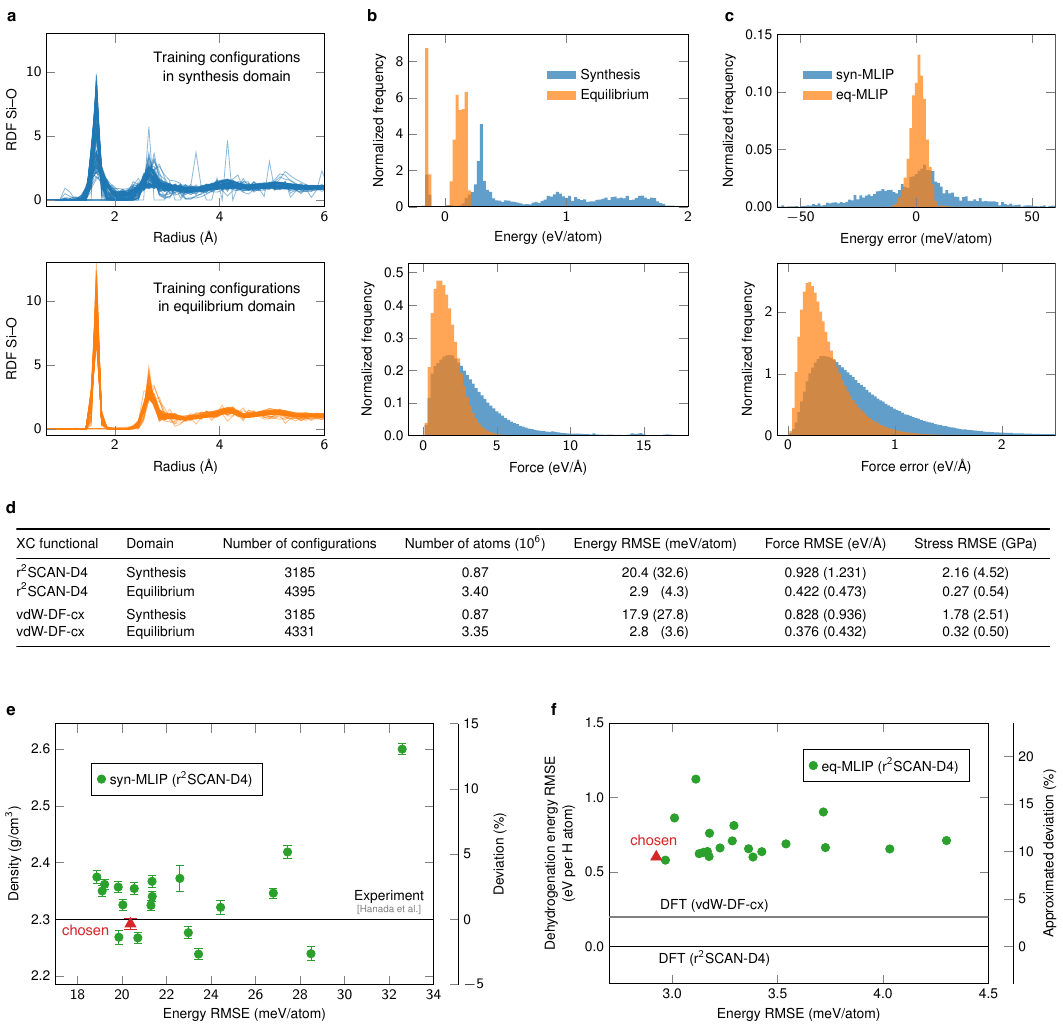}  
   \caption{
   \textbf{Training dataset features and lightweight MLIP training.}
   \textbf{a}, Radial distribution function (RDF) of Si--O pairs from 100 randomly selected training configurations in the synthesis and equilibrium domains.
   Configurations from the synthesis domain exhibit greater variation, reflecting a broader structural diversity.
   \textbf{b},~Distribution of energies and force norms of all configurations in the training datasets calculated using density functional theory (DFT) with the \rtwoscan-D4 functional. 
   Energies are referenced to $\alpha$-quartz (\ce{SiO2}), $\alpha$-alumina (\ce{Al2O3}), an \ce{H2O} molecule, and an \ce{O2} molecule in vacuum relaxed at 0 K.
   Configurations in the synthesis domain exhibit a broader range of relative energies and atomic forces.
   \textbf{c}, Distributions of training errors in energy and force norm, given as deviations of MLIP predictions from DFT values. 
   For both energy and force, syn-MLIP error distributions are skewed to higher values, reflecting larger training errors than for eq-MLIP. 
   \textbf{d}, Training root-mean-square errors (RMSEs). 
   Results with two exchange--correlation (XC) functionals are compared. 
    The values inside brackets show the maximum values from the 20 trained MLIP candidates. 
    Eq-MLIPs exhibit substantially lower training errors than syn-MLIPs in energy, force norm, and stress.
   \textbf{e}, Predicted densities of bulk amorphous metallosilicates at \qty{300}{\kelvin} with an Al/Si molar ratio of 0.17, obtained from syn-MLIP candidates trained on the same dataset.
   Error bars represent 95\% confidence intervals of the mean from multiple synthesis runs. 
   The highlighted syn-MLIP with density closest to the experimental value (Hanada~et~al.~\cite{hanada1989}) and having a relatively low RMSE is used for production runs.
   \textbf{f}, RMSE in dehydrogenation energies at \qty{0}{\kelvin} for eq-MLIP candidates fitted to the same training set. 
   Dehydrogenation is modeled in metallosilicates \ce{SiO2(Al2O3)_{0.1}} with a \qty{1.5}{nm} diameter pore and surface functionality. 
   The highlighted eq-MLIP exhibiting a low RMSE in both training and dehydrogenation energies is chosen for production runs; it achieves an average relative deviation of $\sim$\qty{10}{\percent} in the dehydrogenation energy.
   Good agreement is observed between dehydrogenation energies calculated using DFT with the vdW-DF-cx and \rtwoscan-D4 functionals.
   }
   \label{fig:error}
\end{figure*}

Realistic metallosilicate atomic structures are obtained via MD simulations accelerated by a synthesis-stage potential (\emph{syn-MLIP}), trained on a broad configuration space (Fig.~\ref{fig:workflow}a).
The sampling of metallosilicate environments in the configuration space starts from four regions: 
(i)~crystalline and amorphous SiO$_2$ bulk phases; 
(ii)~chemical reactions starting from a set of molecular SiO$_2$, AlO, and AlO$_2$ units placed in the simulation cell; 
(iii)~melt--quenching of SiO$_2$ and silica--alumina mixtures up to \qty{6000}{\kelvin} to generate amorphous structures; and 
(iv)~additional \textit{ab initio} MD for amorphous and crystalline Al$_2$O$_3$ to capture configurations associated with Al incorporation. 
To describe mesopore environments and water chemistry, the training set also includes configurations containing liquid water, --OH functionalized pore surfaces, and water-filled mesopores.
During MLIP-based MD, physics-informed active learning~\cite{xu24} is employed to automatically expand the training set in poorly sampled regions of configuration space.
In total, the syn-MLIP training set consists of about \qty{3200}{} configurations.

Starting from precursors, we resolve the experimentally synthesized amorphous metallosilicates using an \textit{in silico} synthesis protocol in analogy to experimental processing (Fig.~\ref{fig:workflow}b).
Precursor molecules SiO$_2$ and Al$_2$O$_3$ are placed in the simulation cell according to the target Al/Si composition to generate the amorphous metallosilicate matrix.
Ordered mesopores are introduced by applying a repulsive Lennard--Jones potential placed in the form of a cylinder with its axis running throughout the cell along the $z$-axis~\cite{feuston94}, preventing atoms from occupying the pore region and thereby defining the pore size by the potential cutoff.
The pores can adopt different geometrical arrangements depending on the applied periodic boundary conditions. 
This approach aligns with experimental synthesis~\cite{itzigehl25, bruckner21, meynen09, attard95}, where precursor molecules, e.g.,~\ce{Si(OCH3)4} and \ce{Al(NO3)3}, are mixed with aqueous surfactant solutions.
Driven by the hydrophobic effect, the surfactant forms cylindrical micelles that exclude the precursor molecules from the micellar domain. 
In the simulated synthesis, a melt--quench MD procedure of \qty{1}{ns} is then employed to synthesize the amorphous matrix with controlled mesoporosity.
Subsequently, a functionalization step is performed by introducing --OH groups onto the pore surfaces, followed by annealing using syn-MLIP.
Finally, the system is equilibrated at \qty{300}{\kelvin}, and H$_2$O molecules formed from unstable --OH groups are removed to yield a stable mesoporous metallosilicate model with surface functionality.

Since syn-MLIP is trained to cover a broad configuration space, its accuracy near equilibrium is inherently limited.
To obtain a high-accuracy potential for equilibrium and near-equilibrium structures, we generate a new training set using syn-MLIP-based MD sampling of the targeted systems, i.e., amorphous and hydrated porous metallosilicates (Fig.~\ref{fig:workflow}c), at relatively low temperatures up to \qty{500}{\kelvin}.
The sampling is enriched via active learning with a reduced extrapolation threshold to concentrate on well-relaxed configurations.
The separation of synthesis and equilibrium domains is justified by the high glass transition temperature of metallosilicates, experimentally reported to exceed \qty{1000}{\kelvin}~\cite{Wilke_Benmore_Ilavsky_Youngman_Rezikyan_Carson_Menon_Weber_2022, Liu_Boffa_Yang_Fan_Meng_Yue_2018}, consistent with findings that silica surface roughness is primarily determined during \textit{in silico} quenching~\cite{Wimalasiri_Nguyen_Senanayake_Laird_Thompson_2021}.  
Hydrated metallosilicate surfaces are expected to remain stable~\cite{Gierada_Petit_Handzlik_Tielens_2016, Zhuravlev_2000} or exist in dynamic equilibrium~\cite{Heard_Grajciar_Rice_Pugh_Nachtigall_Ashbrook_Morris_2019} under \textit{operando} conditions near room temperature. 
We additionally include flat surfaces represented by slab models to the training set, improving the stability of the potential without significantly increasing the training error.
In total, about \qty{4400}{} configurations form the training dataset of the equilibrium-stage potential, \emph{eq-MLIP}.

MD simulations on long time scales become accessible with eq-MLIP, yielding predictions that are quantitatively comparable to experimental measurements (Fig.~\ref{fig:workflow}d).
Mesoporous metallosilicates with amorphous matrices are finally synthesized and characterized experimentally, providing a direct basis for validating the end-to-end framework and confirming the accuracy of the resulting MLIPs (Fig.~\ref{fig:workflow}e).

\subsection*{Training lightweight MLIPs}

To highlight the differences in the local atomic environments of the syn-MLIP and eq-MLIP training datasets, we analyze 100 configurations randomly selected from each dataset. 
The respective radial distribution functions of Si--O pairs are shown in Fig.~\ref{fig:error}a. 
Compared to the equilibrium-state configurations, the dataset from the synthesis process exhibits greater variation in Si--O bond lengths, reflecting more diverse local environments. 
This is further emphasized in the energies and atomic forces of the configurations labeled by DFT calculations using the \rtwoscan-D4 functional (Fig.~\ref{fig:error}b): The synthesis dataset includes configurations with formation energies up to \qty{1.8}{eV/atom}, whereas equilibrium configurations reach only \qty{0.3}{eV/atom}. 
Correspondingly, forces as large as \qty{15}{eV/\angstrom} appear in the synthesis dataset, while those in the equilibrium dataset remain within \qty{5}{eV/\angstrom}. 
These results clarify that the melt--quench and functionalization processes sample many high-energy configurations, whereas predicting equilibrium material properties near room temperature requires only low-energy configurations. 

Following the domain-specific training strategy, the two datasets generated from the synthesis process and from equilibrium-state sampling are used to train syn-MLIP and eq-MLIP, respectively. 
Both MLIPs are trained with the same model complexity (608 fitting coefficients) but tailored to different domains of phase space and application.
The training-error distributions are shown in Fig.~\ref{fig:error}c. 
Syn-MLIP exhibits higher fitting errors, as large as \qty{\pm50}{meV/atom} in energies and more than \qty{2}{eV/\angstrom} in forces, whereas eq-MLIP shows much lower errors, mostly within \qty{\pm10}{meV/atom} and \qty{1}{eV/\angstrom} in energies and forces, respectively.
Figure~\ref{fig:error}d summarizes average training errors for lightweight MLIPs separately trained on datasets labeled by two dispersion-aware functionals: r$^2$SCAN with the empirical D4 dispersion correction and the nonlocal vdW-DF-cx functional.
%
%
%
%
The energy root-mean-square errors (RMSEs) of syn-MLIP (\qtyrange{17}{21}{meV/atom}) are consistent with previous studies on amorphous systems, which report values from \qty{10}{meV/atom}~\cite{erlebach2024} to \qty{36}{meV/atom}~\cite{leimeroth2024}, but remain larger than typical errors for crystalline systems (\qty{2}{meV/atom})~\cite{forslund23,jung2023,zhou2022}.
In contrast, eq-MLIPs achieve substantially lower errors---about fivefold lower in energy (\qty{3}{meV/atom})---reaching values comparable to crystalline systems.
These results demonstrate that the domain-specific training strategy enhances the lightweight MLIP accuracy in the relevant part of phase space without increasing the model complexity.

To investigate the MLIP parameter space, we analyze multiple \emph{independent} training runs, each using identical datasets but different random seeds, thereby generating an ensemble of MLIP candidates.
The quality of syn-MLIP candidates is evaluated based on the bulk density of bulk amorphous metallosilicates (Fig.~\ref{fig:error}e), which constitutes a critical material property.
The training errors in energy ranging from \qtyrange{19}{33}{meV/atom} result in uncertainties of \qtyrange{2.2}{2.6}{g/cm^3} in the predicted density, which corresponds to errors of up to \qty{13}{\percent} relative to the experimental value (about \qty{2.3}{g/cm^3})~\cite{hanada1989}.
Within this ensemble of 20 candidates, the MLIP with the lowest training error does not yield a density in close agreement with experiment. 
Therefore, the MLIP reproducing best the experimental density and having a relatively small training error is chosen to validate the end-to-end framework.

\begin{figure*}[tbh!]  
   \centering  %
    \includegraphics{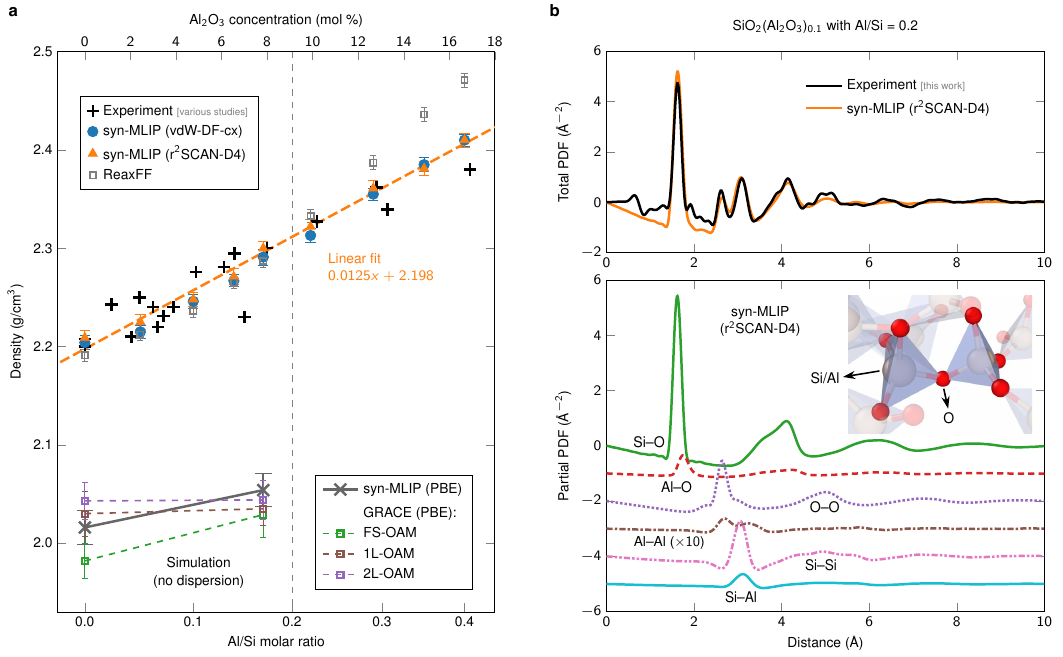}  
   \caption{
   \textbf{Validation of syn-MLIP on bulk amorphous metallosilicates.}
   \textbf{a}, Bulk densities of \ce{SiO2(Al2O3)_{$x$/2}} at \qty{300}{\kelvin} versus Al/Si molar ratio $x$. 
   Predictions from both syn-MLIPs, trained separately on vdW-DF-cx and r$^2$SCAN-D4 data, agree well with experimental results from various studies~\cite{ando2018,morikawa1982,khemis2024,hanada1989,ohira2016}. 
   A linear relationship between density and Al/Si ratios up to 0.4 is confirmed by a fit to the simulation data, with coefficients of determination of $R^2 = 0.99$ for r$^2$SCAN-D4. 
   ReaxFF~\cite{zhang24} overestimates the density containing more than \qty{12}{mol.\%} \ce{Al2O3}.
   GRACE models~\cite{PhysRevX.14.021036} referenced to PBE underestimate the density by about 10\%.
   For comparison, a syn-MLIP referenced to PBE shows results consistent with the GRACE models. 
   Error bars represent 95\% confidence intervals of the mean from multiple synthesis runs. 
   \textbf{b}, Simulated and experimental pair distribution functions (PDFs) at an Al/Si molar ratio of 0.2 and \qty{300}{\kelvin}. 
   The atomistic model synthesized \textit{in silico} with syn-MLIP quantitatively reproduces experimental total PDFs. 
   Partial PDFs derived from the atomistic model are compared, providing insight for interpreting experimental data. 
   }
   \label{fig:validation}
\end{figure*}

\begin{figure*}[tbh!]  
   \centering  %
    \includegraphics{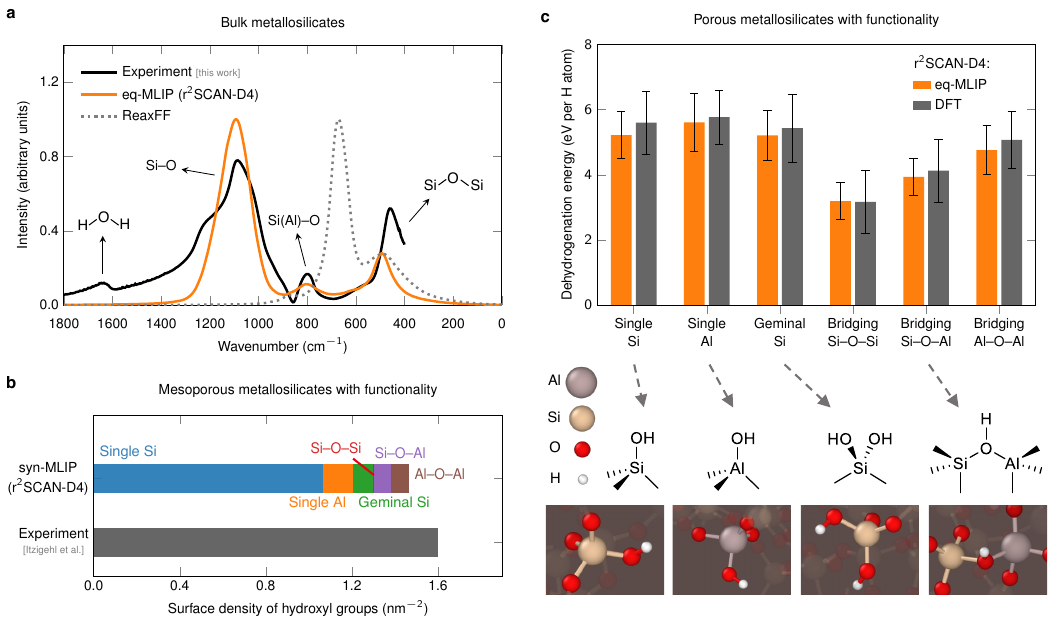}  
   \caption{
   \textbf{Validation of eq-MLIP and of the complete end-to-end framework.}
   \textbf{a}, 
   Comparison of simulated and experimentally measured infrared spectra of bulk amorphous metallosilicates \ce{SiO2(Al2O3)_{0.025}}.
   The end-to-end framework is employed: \rtwoscan-D4-based syn- and eq-MLIPs are used for \textit{de novo} \textit{in silico} synthesis and spectrum calculation, respectively, whereas ReaxFF~\cite{zhang24} is used for both tasks. 
   Good agreement for the Si--O peak confirms the accuracy of both the r$^2$SCAN-D4 functional and eq-MLIP.
   The limitation of ReaxFF is also evident. 
   \textbf{b}, Surface densities of hydroxyls determined from the atomistic model synthesized \textit{in silico} and from our previous experiments~\cite{itzigehl25}.
   Both simulation and experiment examine ordered mesoporous metallosilicates \ce{SiO2(Al2O3)_{0.1}} with \SI{6.3}{nm} diameter pores, forming a two-dimensional hexagonal lattice with a lattice constant of \SI{10.1}{nm}, and with surface functionality.
   Experimental measurements based on reactions of surface hydroxyls with a Grignard reagent quantify all surface-accessible  hydroxyls~\cite{itzigehl25}, whereas different hydroxyl species are resolved from the atomistic model via their local atomic environments.
   The 95\% confidence intervals of the mean predicted densities for all hydroxyl types are below \qty{0.03}{nm^{-2}}.
   The small discrepancy in total density (about 10\%) supports the realism of the atomistic model obtained from the end-to-end framework.
   \textbf{c},~
   Dehydrogenation energies at \qty{0}{\kelvin} predicted by eq-MLIP and DFT.
   Metallosilicates \ce{SiO2(Al2O3)_{0.1}} have a pore diameter of \qty{1.5}{nm} and are synthesized and functionalized \textit{in silico} using syn-MLIP. 
   Dehydrogenation is modeled on the pore surface with eq-MLIP, and error bars indicate the standard deviation arising from different hydroxyl sites on the pore surface. 
   Predicted dehydrogenation energies are similar for the single and geminal hydroxyls and generally lower for bridging hydroxyls.
   The agreement demonstrates that eq-MLIP maintains near-\textit{ab initio} accuracy.
   Illustrative structural formulas are shown, and atomic sites for dehydrogenation are visualized.
   }
   \label{fig:eq_validation}
\end{figure*}

Dehydrogenation is an important chemical process related to surface acidity, yet resolving the energetics of this process at different atomic sites is experimentally challenging~\cite{Busca2019, Busca2017, Raybaud2011, Deng2006}. 
We therefore examine the quality of eq-MLIP candidates in predicting dehydrogenation energies against DFT (Fig.~\ref{fig:error}f). 
Eq-MLIP training errors of \qtyrange{2.9}{4.3}{meV/atom} lead to errors of \qtyrange{0.5}{1.1}{eV.per.H.atom} in the dehydrogenation energy, corresponding to approximately \qtyrange{9}{17}{\percent} deviation.
Dehydrogenation energies derived from ReaxFF~\cite{zhang24} exhibit a comparable error of \qty{0.78}{eV.per.H.atom} relative to the PBE-D3 DFT baseline.
The relatively large deviations of MLIP dehydrogenation energies (about 10\%) stem from the high atom count of simulation models ($\sim$\qty{700}{atoms}) necessary for constructing the pores.
From the eq-MLIP ensemble, we select the MLIP with the lowest errors in both training and dehydrogenation-energy predictions for the following framework validation. 
%

%

\begin{figure*}[t]  
   \centering  %
    \includegraphics{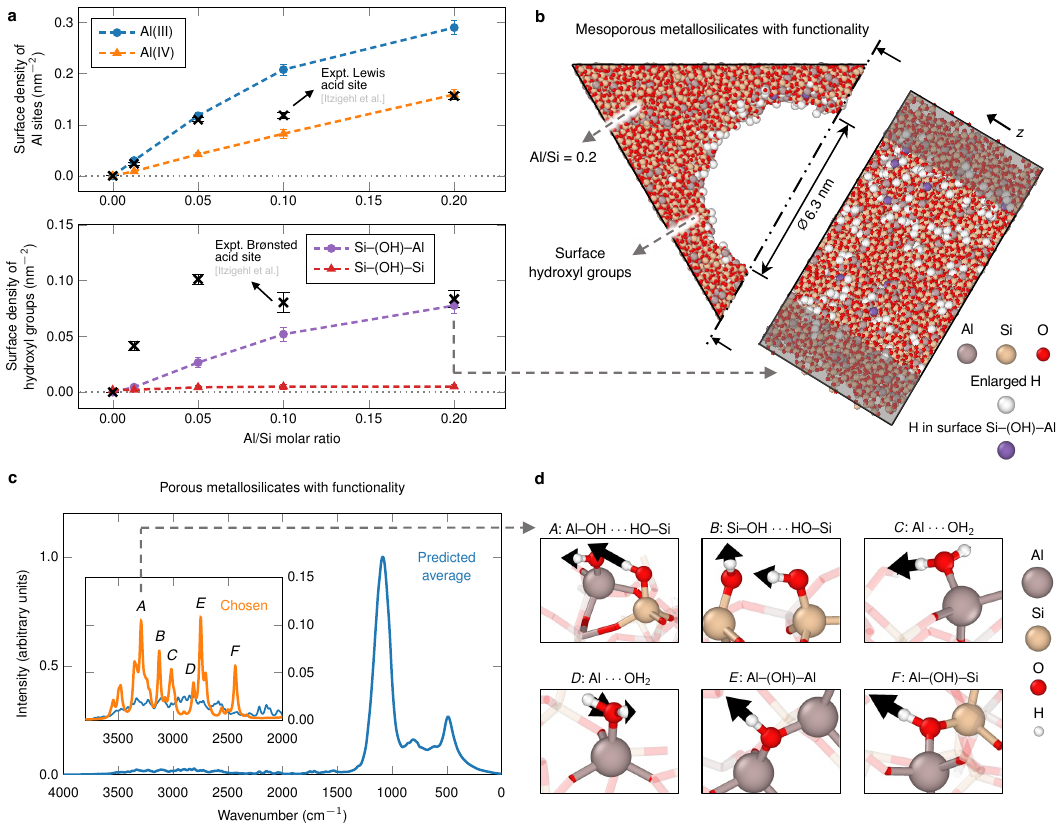}  
   \caption{
   \textbf{Example applications of the end-to-end framework for metallosilicates under \textit{operando} conditions.}
   \textbf{a},~Surface density of acid sites versus Al/Si molar ratio from atomistic models and our previous experimental measurements~\cite{itzigehl25}. 
   Predictions are based on three- and four-fold coordinated Al sites for Lewis acidity and bridging hydroxyls for Brønsted acidity.
   Error bars represent 95\% confidence intervals of the mean from multiple synthesis runs. 
   Experimental values are obtained from Fourier-transform infrared spectroscopy after pyridine loading and desorption at \qty{473}{\kelvin}~\cite{itzigehl25}. 
   Simulated structures exhibit pore diameters and hexagonal lattice parameters comparable to experimentally synthesized materials.
   \textbf{b},~Atomistic model of ordered mesoporous \ce{SiO2(Al2O3)_{0.1}} with surface functionality. 
   Enlarged H atoms indicate the location of hydroxyl groups, with bridging Si--(OH)--Al groups at the pore surface specifically highlighted in purple.
   Undercoordinated Al sites are also present at the pore surface.
   \textbf{c},~Predicted infrared spectra of metallosilicates \ce{SiO2(Al2O3)_{0.1}} with \qty{1.5}{nm} diameter pores and surface functionality.
   Spectra averaged over ten MD snapshots are shown to mimic experimental conditions.
   One chosen snapshot is compared in the inset with an enlarged intensity scale. 
   \textbf{d},~Six vibrational modes related to --OH functional groups are resolved, with their displacement patterns visualized by black arrows in the atomistic models.
   MLIPs referenced to the \rtwoscan-D4 functional are used. 
   }
   \label{fig:prediction}
\end{figure*}

\subsection*{Experimental validation}

%
We experimentally synthesize and characterize bulk and mesoporous metallosilicates in parallel to atomistic modeling.
%
Observations from experiments are then compared with results extracted from large-scale MD simulations accelerated by eq-MLIP, based on the atomic structures synthesized \textit{in silico} by syn-MLIP.

First, \textit{in silico}-synthesized bulk metallosilicates with varying compositions are validated against experimental measurements reported in multiple studies (Fig.~\ref{fig:validation}a). 
The predicted bulk densities for both functionals---vdW-DF-cx and r$^2$SCAN-D4---are consistent with each other and quantitatively reproduce experimental values for metallosilicates with Al incorporation up to an Al/Si molar ratio of 0.4.
Our high accuracy simulation results further confirm the linear relationship between metallosilicate density and \ce{Al2O3} concentration in the investigated regime, as hypothesized in experimental studies~\cite{aksay1979,khemis2024}.
By comparison, ReaxFF~\cite{zhang24} (following the same \textit{de novo} \textit{in silico} synthesis protocol) overestimates the density for $\textrm{Al/Si} \geq 0.3$ (cf.~Supplementary Fig.~S12). 
%
In strong contrast, GRACE models systematically underestimate the density by as much as 10\% (\qty{2.0}{} vs.~\qty{2.2}{g/cm^{3}} for silica).
An additional syn-MLIP trained on PBE DFT data yields densities consistent with the GRACE predictions. 
These results indicate that functionals incorporating explicit vdW interactions are necessary to capture metallosilicate densities within DFT, consistent with a recent study~\cite{Pedone_Bertani_Benassi_2025}. 

While density is an averaged property, we further validate syn-MLIP for local atomic environments by comparing predicted PDFs with experimental data measured in the present study.
Experimental PDFs are obtained from total scattering measurements on bulk metallosilicate powders with $\mathrm{Al/Si} = 0.2$.
In parallel, simulated PDFs are extracted from MD simulations with syn-MLIP at \qty{300}{\kelvin} using the \textit{in silico}-synthesized atomistic model with the same Al/Si ratio. 
The simulations accurately reproduce the peaks in the total PDFs observed experimentally (Fig.~\ref{fig:validation}b), demonstrating that the obtained atomistic models capture the local structure of experimentally synthesized samples. 
These results confirm the high accuracy of syn-MLIP and support the validity of the utilized \textit{de novo} \textit{in silico} synthesis protocol. 
Furthermore, decomposition of the simulated PDFs provides element-resolved structural information that cannot be reliably obtained from experiments~\cite{weber2008, okuno2005}.
The assigned peaks are in good agreement with previously reported positions in aluminosilicate glasses~\cite{wilding2010, bernasconi2021, Wilke_Benmore_Ilavsky_Youngman_Rezikyan_Carson_Menon_Weber_2022, urata2021, ohkubo2024}.

Vibrational properties provide quantitative validation of the melt--quench--synthesized structures and the near-equilibrium potential-energy surface described by eq-MLIP.
Experimentally, we obtain the vibrational spectrum of bulk amorphous metallosilicates \ce{SiO2(Al2O3)_{0.025}} from Fourier-transform infrared spectroscopy. 
Under the end-to-end framework, MD simulations with eq-MLIP are performed on atomistic models synthesized by syn-MLIP to extract an effective harmonic potential, from which infrared spectra are obtained via polarization calculations using DFT. 
Figure~\ref{fig:eq_validation}a shows that infrared spectra predicted by the end-to-end framework accurately reproduce the main vibrational bands at \qty{1100}{cm^{-1}} (Si--O antisymmetric stretching) and \qty{400}{cm^{-1}} (Si--O--Si bending)~\cite{okuno2005}, confirming the high accuracy of the end-to-end framework accelerated by the lightweight MLIPs.
By contrast, applying the same end-to-end protocol with the ReaxFF potential~\cite{zhang24} yields infrared spectra that significantly deviate from experimental measurements. 
Similar limitations of ReaxFF have also been reported in comparisons with neutron scattering functions~\cite{galaviz25}.
Discrepancies between experiments and simulations above \qty{1200}{cm^{-1}} are attributed to residual water in the experimental samples, as evidenced by the characteristic \ce{H2O} scissoring band near \qty{1600}{cm^{-1}}~\cite{Palencia_2018}.

%
For functionality validation, we generate ordered mesoporous metallosilicates with geometric parameters closely matching experimental samples, including the pore size (\qty{6.3}{nm}) and the hexagonal lattice parameter (\qty{10.1}{nm}). 
The resulting surface hydroxyl densities agree with experimental values determined via reaction with a Grignard reagent (\qty{1.47}{} vs.~\qty{1.60}{nm^{-2}}, Fig.~\ref{fig:eq_validation}b), with an underestimation of about 10\%, well within the experimentally observed variations arising from different pretreatment~\cite{vanderVoort.1991} or calcination~\cite{Benamor.2012} conditions.
%

Finally, we validate the lightweight MLIP’s accuracy in capturing dehydrogenation energetics at metallosilicate pore surfaces (Fig.~\ref{fig:eq_validation}c).
For each dehydrogenation type, over 37 distinct surface sites are sampled to ensure robust statistics.
The close agreement between eq-MLIP and DFT demonstrates the near--\textit{ab initio} accuracy of eq-MLIP, in line with its low fitting RMSEs.
Among the examined environments, bridging Si--OH--Si and Si--OH--Al hydroxyl groups exhibit a low dehydrogenation energy of \qtyrange{3.5}{4}{eV.per.H.atom}, indicating weak hydrogen binding.
Dehydrogenation of single and geminal silanol groups requires \qtyrange{5}{6}{eV.per.H.atom}, in agreement with the O--H bond dissociation energy of silanol (about \qty{5.2}{eV.per.H.atom})~\cite{Lucas_Curtiss_Pople_1993}.

\subsection*{Example applications}

Mesoporous confinement strongly influences catalytic behavior as demonstrated for metathesis polymerization~\cite{probst2025}. Strength and density of acid sites likewise govern catalytic performance~\cite{Secci.2023}.
Atomistic modeling can guide \textit{a priori} design of mesoporous metallosilicates for catalytic processes, but requires realistic structures that resolve surface acid sites across compositions.
Such structures can be obtained using our end-to-end framework, thereby going beyond previous studies that focused primarily on pure silica~\cite{Rubirigi_Aprile_Champagne_2026, P.Klein_A.Pidko_A.Kolganov_2025, ewing14}.
Densities of Al sites and bridging hydroxyls at the pore surfaces of metallosilicates with varying Al loadings are shown in Fig.~\ref{fig:prediction}a. 
At low Al concentrations of $\textrm{Al/Si} < 0.1$, experimental Lewis acid site densities closely track under-coordinated Al(III) sites, whereas at higher loadings such as $\textrm{Al/Si} = 0.2$, they are instead close to the predicted Al(IV) densities. This suggests an increased contribution of Al(IV) sites to the acidic environments in experimental samples with increasing Al incorporation, consistent with experimental observations~\cite{itzigehl25}.
The Si--(OH)--Al groups reach a surface density of \qty{0.078}{nm^{-2}} at $\textrm{Al/Si} = 0.2$ (Fig.~\ref{fig:prediction}b), in close agreement with experimentally measured Brønsted acid-site densities (\qty{0.084}{nm^{-2}}).
Bridging Si--(OH)--Si densities remain below \qty{0.01}{nm^{-2}}, indicating a negligible contribution to Brønsted acidity.
Discrepancies may be resolved by accounting for pseudo-bridging groups~\cite{larmier17, Raybaud2011} and acidity strength in the predictions. 
Embedding catalysts into such realistic mesoporous metallosilicates with functionality (Extended Data Fig.~\ref{fig:catalyst}) enables probing the impact of surface acid sites on catalytic performance and accelerates targeted optimization of metallosilicate compositions.

Adsorption at metallosilicate acid sites is another key application, with direct relevance to industrial processes~\cite{Jiang_Zhong_Jafari_Du_He_Fu_Singh_Suib_2016, El-Safty_Shahat_Ismael_2012, Li_Liu_Xing_Gao_Xiong_Li_Li_Liu_2007}. 
Experimentally, the nature of acid sites can only be probed indirectly, for example via pyridine desorption~\cite{itzigehl25}. 
Corresponding results are strongly influenced by spectral averaging and \textit{operando} conditions~\cite{Coumans.2024, Yang.2021c,goldsmith2017}, leaving the local structure of acid sites heavily debated~\cite{Wang.2021, Salvia.2023, Lafon.2025}. 
We first mimic experimental conditions by averaging infrared spectra from ten MD snapshots (Fig.~\ref{fig:prediction}c). 
Apart from the Si/Al--O amorphous matrix bands below \qty{1500}{cm^{-1}} (cf.~Fig.~\ref{fig:eq_validation}a), the high-wavenumber region (\qtyrange{2000}{3700}{cm^{-1}}) displays broad, weak features.
Analysis of a single snapshot resolves these features into distinct peaks, assignable to hydrogen vibrations in hydroxyls adjacent to Si or Al atoms (Fig.~\ref{fig:prediction}d).
At higher wavenumbers, vibrations arise from hydroxyls bound to Si or Al sites (\textit{A} and \textit{B}), with additional contributions from water molecules coordinated to surface Al sites (\textit{C} and \textit{D}).
The predicted spectrum also shows that bridging Al--(OH)--Si groups (\textit{F}) are associated with lower-wavenumber bands, indicating weaker O--H bond strength, consistent with previous studies~\cite{Raybaud2011,perez-beltran16,treps2021}.
These atomistic insights from realistic structures help understand adsorption phenomena, both simulated (Extended Data Fig.~\ref{fig:acidity}) and observed experimentally~\cite{itzigehl25, sousa2025}, enabling the rational design of mesoporous adsorbents with tailored surfaces.

\section*{Discussion}

\begin{figure}[t]  
   \centering  
    \includegraphics{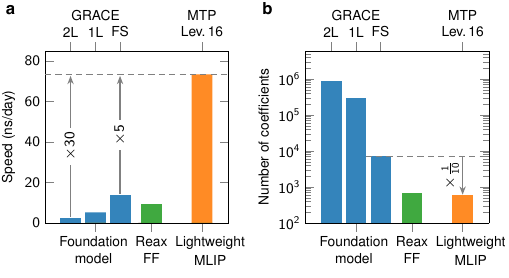}
   \caption{
   \textbf{Lightweight MLIP advantages.}
   \textbf{a}, MD simulation speed. 
   Performance is evaluated on a 700-atom model using a single NVIDIA V100S GPU with lightweight moment tensor potentials (MTPs) at level 16, GRACE foundation models~\cite{PhysRevX.14.021036} at different levels of complexity and ReaxFF~\cite{zhang24}. 
   The lightweight MLIP is more than five times faster than the other potentials. 
   \textbf{b}, Number of fitting coefficients on a logarithmic scale. 
   Only coefficients related to the elements Si, Al, H, and O are counted. 
   The lightweight MLIP contains a number of fitting coefficients comparable to ReaxFF and more than an order of magnitude fewer than foundation models. 
   }
   \label{fig:cost}
\end{figure}

High computational cost and the lack of reliable simulation protocols have traditionally limited atomistic modeling, preventing its application to intrinsically complex materials such as amorphous metallosilicates.
Our end-to-end computational framework has enhanced the capabilities of atomistic modeling by leveraging efficient lightweight MLIPs (Fig.~\ref{fig:cost}) trained with a domain-specific strategy and by closely mimicking experimental synthesis. 
The resulting simulations reproduce multiple experimental observables with quantitative agreement, providing confidence in resolving realistic structures beyond the reach of previous studies~\cite{strugovshchikov26, Roy_Dürholt_Asche_Zipoli_Gómez-Bombarelli_2024}.
Our tools for simulating experimental observables are open source, motivating experimental validation within the community and providing the basis for standardized workflows to evaluate MLIPs.

Further application of our end-to-end framework enables medium-throughput screening of metallosilicates across diverse metal species, loadings, and pore geometries.
For example, systems containing Y or Ti can be studied by replacing the precursors with \ce{Y2O3} or \ce{TiO2} prior to melt--quench simulations, while reusing existing DFT data for common components such as pure silica and water. 
Datasets generated using different exchange--correlation functionals provide a basis for extending the framework beyond DFT accuracy~\cite{Ikeda_2026}.
Alongside chemical composition, pore architecture can be tuned via modification of the repulsive potentials, enabling the construction of three-dimensional pore networks~\cite{El-Safty_Shahat_Ogawa_Hanaoka_2011, meynen09}. 
Looking ahead, \textit{in silico}-synthesized diverse structural ensembles, annotated with the type, strength, and density of acid sites, provide high-quality training data for composition--structure--property generative models~\cite{Yang_Schwalbe-Koda_2025, Xie_2025}, thereby enabling rapid and targeted discovery of novel metallosilicates for experimental synthesis.

The absence of explicit charge effects in the present work limits the \textit{operando} conditions that can be simulated and compared with experiments.
Incorporating non-local charge-transfer mechanisms into lightweight MLIPs, such as those implemented in fourth-generation neural network potentials~\cite{Ko_Finkler_Goedecker_Behler_2021}, provides a natural pathway to overcome this limitation.
Additionally, coupling our framework with emerging foundation models may accelerate training data generation~\cite{Li_Charoenphakdee_Zhuang_Okuno_Tsuboi_Takamoto_Ishida_Li_2025} and broaden its applicability to more complex reactive chemical processes.



As experimental synthesis techniques advance, atomistic modeling will play an increasingly central role in guiding the design of next-generation functional materials, from the metallosilicates studied here to, e.g., solid electrolytes for battery~\cite{Ou2024} or special alloys for hydrogen storage~\cite{Kumar_2026}.
Our simulation--experiment collaborative work represents a step toward addressing the long-standing challenge of improving the realism and reliability of atomistic modeling.

\onecolumngrid


\clearpage
\twocolumngrid

\section*{METHODS\label{sec:methods}}


\subsection*{Simulations}

\subsubsection*{Moment tensor potentials}

Moment tensor potentials (MTPs) at level 16 with four chemical species (Al, Si, O, and H; 608 fitting parameters) and a cutoff distance of \qty{5}{\angstrom} were fitted using the \textsc{mlip-}\scalebox{0.8}{3} code~\cite{shapeev16,podryabinkin23}. 
The fitting weights for energy, force, and stress were set to $1/N_i$, \qty{0.01}{\angstrom\squared}, and $0.001/N_i$, respectively, where $N_i$ denotes the number of atoms in the $i$-th configuration. 
Tests show that increasing the cutoff to \qty{6}{\angstrom} does not reduce the training errors.
Active learning~\cite{novikov21} was performed in a physics-informed manner~\cite{xu24} to optimize syn-MLIP and eq-MLIP for their target applications. 
Various compositions and structures were sampled, including slab models and structures with \qty{1.5}{nm}-diameter pores (containing more than 1200 atoms), to incorporate information on the extreme limits of surface curvature into the training data. 
In active learning, molecular dynamics (MD) protocols were tailored to ensure stable simulations of target processes, e.g., melt--quench. 
MTP-based MD simulations were performed using \textsc{lammps}~\cite{plimpton95,thompson22}, optionally with a \textsc{Kokkos} implementation for GPU acceleration~\cite{meng25}. 
Details of the sampling process for the training structures and the training of the MTPs are provided in Supplementary Secs.~S1 
 and~S2, 
 respectively.

\subsubsection*{Electronic-structure calculations}

Electronic-structure calculations were performed under the density functional theory (DFT) framework with the projector augmented-wave (PAW) method~\cite{blochl94,kresse99} implemented in \textsc{vasp}~\cite{kresse96a}. The valence-electron configurations in the PAW potentials were: Si: $3s^2 3p^2$; O: $2s^2 2p^4$; Al: $3s^2 3p^1$; H: $1s^1$. For the exchange--correlation functional, a generalized gradient approximation (GGA) with non-local van der Waals (vdW) correction, i.e., \dfcx \cite{berland14}, and a meta-GGA functional with a semi-classical vdW correction, i.e., \rtwoscan-D4 \cite{furness20,ehlert21}, were used.

High-accuracy DFT calculations were performed for the training structures. Energy cutoffs of \qty{520}{eV} for \dfcx\ and \qty{600}{eV} for \rtwoscan-D4 were used, corresponding to 1.3$\times$\texttt{ENMAX} and 1.5$\times$\texttt{ENMAX}, respectively (with \texttt{ENMAX} the default plane wave cutoff for the PAW potentials). 
A $\mathbf{k}$-point spacing (\texttt{KSPACING} flag) of \qty{0.52}{\per\angstrom} was used with a Gaussian smearing of \qty{0.1}{eV}. 
For training-set generation via DFT MD and dehydrogenation energy calculations via DFT, an energy cutoff of \qty{400}{eV} ($\texttt{ENMAX}$) was tested and found to be sufficient, and was therefore adopted. Active learning during the sampling of training structures was performed using the \dfcx functional.

\subsubsection*{Synthesis of bulk metallosilicates}

Bulk amorphous metallosilicates with a composition of \ce{SiO2(Al2O3)}$_{x/2}$ and varying Al/Si molar ratios $x$ were generated in two steps. 
First, \ce{SiO2} and \ce{Al2O3} precursor molecules were randomly inserted into a cubic simulation box to generate an initial structure matching the experimental density of binary bulk amorphous metallosilicates with the target Al/Si molar ratio~\cite{ando2018,morikawa1982,khemis2024,hanada1989,ohira2016}. 
Subsequently, the resulting structure was subjected to a melt--quench process implemented in MD simulations~\cite{Yang2019,erhard2022,Nakano1994,pedone2009,kayano24}, consisting of \textit{NVT} heating from 300 to \qty{5000}{K} at \qty{23.5}{K/ps}, \textit{NVT} melting at \qty{5000}{K} for \qty{200}{ps}, \textit{NVT} quenching from 5000 to \qty{4000}{K} at \qty{10}{K/ps}, \textit{NpT} quenching from 4000 to \qty{300}{K} at \qty{5}{K/ps}, and final equilibration at \qty{300}{K} for \qty{250}{ps}. The Nos\'e-Hoover thermostat was used for \textit{NVT} and the Nos\'e-Hoover barostat for \textit{NpT}, with a timestep of \SI{1}{fs}~\cite{shinoda04}.


\subsubsection*{Synthesis of porous metallosilicates with surface functionality}
Porous amorphous metallosilicates with a composition of \ce{SiO2(Al2O3)}$_{x/2}$ were generated analogously to the bulk metallosilicates but employing a hexagonal simulation cell.
To generate the pore, a repulsive Lennard--Jones potential (see Supplementary Sec.~S3 A
) was applied in the form of a cylinder with its axis running throughout the simulation cell along the $z$-axis, following previous studies~\cite{feuston94,Han_Slowing_Evans_2020}.  
 
%

Surface functionality was introduced to the generated porous metallosilicates via formal hydroxylation~\cite{wimalasiri21} of bridging $X$--O--$X'$ groups ($X$, $X' =$ Si, Al). 
%
%
Specifically, undercoordinated three-fold Si and dangling O atoms at the pore surface were first saturated, followed by hydroxylation of all surface-accessible bridging $X$–O–$X'$ groups to form $X$–OH and $X'$–OH terminations.
Surface-accessibility of the $X$--O--$X'$ groups was determined using \textsc{pyzeo}~\cite{willems12,pyzeo25}, with a probe radius of \qty{1.4}{\angstrom}, close to that of a \ce{H2O} molecule\cite{willems12,decherchi2015,zefirov1989}. 
The resulting structures were relaxed and subjected to an annealing process in the \textit{NpT} ensemble using the Nos\'e-Hoover barostat with a timestep of \SI{1}{fs}~\cite{shinoda04}, comprising heating from 300 to \qty{820}{K} at \qty{5}{K/ps}, annealing at \qty{820}{K} for \qty{2}{ns}, quenching from 820 to \qty{300}{K} at \qty{5}{K/ps}, and equilibration at \qty{300}{K} for \qty{300}{ps}. 
Finally, \ce{H2O} and \ce{H3O+} molecules formed via dehydroxylation of $X$--OH groups during relaxation or annealing were removed. 


\subsubsection*{Bulk densities}
Densities were calculated from bulk metallosilicates containing about 700 atoms synthesized using syn-MLIP, ReaxFF, or GRACE potentials. 
A syn-MLIP referenced to PBE~\cite{Perdew_Burke_Ernzerhof_1996} data  was developed for comparison with the PBE-based GRACE models (see Supplementary Table~S4).
%
For configurations generated with GRACE, the quenching rate was increased to 50~K/ps. 
This modification was found to have a negligible effect on the average densities (see Supplementary Fig.~S4).

%
%
%
The density of each configuration was obtained by time-averaging the instantaneous density over the final 200~ps of the 300~K MD simulation. 
For each Al/Si ratio, the density was averaged over 100 configurations. 

\subsubsection*{Surface densities}
Mesoporous metallosilicates containing about 20\,000 atoms and surface functionality were synthesized using \rtwoscan-D4 syn-MLIP. 
To match experimental samples~\cite{itzigehl25}, initial pore diameters of 6.8, 6.1, 6.3, 6.1, and \qty{6.3}{nm}, and initial lattice parameters of $a =$ 9.9, 9.9, 9.7, 9.9, \qty{10.1}{nm} and $c = \qty{5}{nm}$ were used for Al/Si molar ratios of 0.0, 0.0125, 0.05, 0.1, and 0.2, respectively.
Surface density was calculated as $N_{\textrm{acc}}/A_{\textrm{acc}}$, where $N_{\textrm{acc}}$ denotes the amount of functional groups accessible on the pore surface and $A_{\textrm{acc}}$ the accessible pore surface area. 
Surface accessibility was determined using spherical probes (see \textit{Synthesis of porous metallosilicates with surface functionality}) and averaged over 40 metallosilicate samples. 
A probe radius of 3.26~\AA, comparable to that of a pyridine molecule~\cite{gugeler2025}, was used to identify accessible functional groups, whereas a probe radius of 1.86~\AA, comparable to that of an N$_2$ molecule~\cite{bae2010}, was used to determine the accessible surface area. 
The sensitivity of the predicted surface densities to the probe size is discussed in Supplementary Sec.~S4 C.

\subsubsection*{Pair distribution functions}

Pair distribution functions (PDFs) weighted by X-ray scattering factors were calculated using \textsc{diffpy-cmi}~\cite{juhas13,juhas15}.
%
%
The limited range of scattering vectors in experiments was simulated using $Q_\textrm{max}=50$ \AA\textsuperscript{-1} and applying a Lorch filter~\cite{prill15}.
Peak broadening was described using a small isotropic thermal displacement parameter $U_\textrm{iso, eqiv}=0.001$ \AA\textsuperscript{2} and by averaging over five 3000-atom samples for each Al/Si ratio. 
%
%

\subsubsection*{Dehydrogenation energies}

Dehydrogenation energies were calculated as the reaction energies for removing an H atom from the pore surface of porous metallosilicates with surface functionality,
\begin{equation}
\begin{aligned}
\Delta E = \,\, &  E(\textrm{metallosilicates w/o an H}) + E(\textrm{H}) \\ 
    & - E(\textrm{metallosilicates}),
\end{aligned}
\end{equation}
where $E(\textrm{metallosilicates w/o an H})$ and $E(\textrm{metallosilicates})$ denote the potential energies of metallosilicates containing about 700 atoms after and before H removal, respectively. 
Further, $E(\textrm{H})$ is the potential energy of an isolated H atom in vacuum, calculated with ReaxFF for ReaxFF-related results and with spin-polarized \rtwoscan-D4 DFT for the remaining cases.  
H atom removal at 697 sites across 91 porous metallosilicate samples was analyzed (single Si:~200; single Al:~185; geminal Si:~100; and bridging Si--O--Si:~75, Si--O--Al:~100, Al--O--Al:~37). 

\subsubsection*{Infrared spectra}

Infrared spectra were calculated for bulk and porous metallosilicates (annealed at 800~K for 10~ns) with 200 and 700~atoms, averaged over 30 and 10 samples, respectively. 
Effective harmonic potentials were obtained by fitting force constants with a cutoff radius of 4~\AA{} to 200 uncorrelated snapshots from \textit{NVT} MD at 20~K using a Langevin thermostat with a friction parameter of 0.01~fs$^{-1}$.
The numbers of fitted parameters are 1.81$\times 10^5$ and 5.64$\times 10^5$, and the corresponding test $R^2$ values are 0.98 and 0.96 for the bulk and porous metallosilicates, respectively. 
The fitting was performed using a modified version of \textsc{hiphive}~\cite{eriksson19} with an efficient iterative algorithm for sparse linear equations as implemented in \textsc{scipy}~\cite{paige82}. 
%
%
%
Born effective charges were calculated within density functional perturbation theory~\cite{baroni86,gajdos06} using the \dfcx functional with an energy cutoff of 400~eV and $\Gamma$-point sampling. 
%
%
%
%
%
%
%
Infrared spectra were simulated from the polarization response of vibrational normal modes, using the Born effective charge tensor and a linewidth of 30 cm$^{-1}$ as implemented in \textsc{phonopy-spectroscopy} \cite{skelton17}. 
%

\subsubsection*{ReaxFF and foundation models}
ReaxFF from Zhang~\textit{et~al.}~\cite{zhang24} was used with the \textsc{Kokkos}-based \texttt{reaxff/kk} pair style~\cite{aktulga2012} implemented in \textsc{lammps}~\cite{plimpton95,thompson22}. 
Atomic charges were computed at each timestep using the \texttt{qeq/reaxff} charge equilibration scheme~\cite{aktulga2012}, with a lower cutoff of 0~\AA{}, an upper cutoff (Taper radius) of 10~\AA{}, a convergence tolerance of 10$^{-6}$~e, and a maximum of 400 iterations. 


Graph atomic cluster expansion (GRACE) foundation models, including GRACE-2L-OAM, GRACE-1L-OAM, and GRACE-FS-OAM, implemented in \textsc{gracemaker}~\cite{PhysRevX.14.021036,lysogorskiy25}, were employed. 
The number of coefficients was determined by restricting the model to the chemical species Si, Al, O, and H. 
%







\subsection*{Experiments}

\subsubsection*{Synthesis of bulk and mesoporous metallosilicates}

Metallosilicates were synthesized following our previously reported protocol~\cite{itzigehl25}. 
In short, tetramethyl orthosilicate (TMOS, Sigma-Alrdich, 99.0\%), \ce{Al(NO3)3 . 9H2O} (Arcos Organics, $\geq 98\%$), and 0.1~M HNO$_3$ (Synergy, purum, 65\%, diluted with deionized water) were mixed in a polytetrafluoroethylene flask and prepolymerized at approximately 80~mbar for 5~min while stirring. 
The mixture was then added to a surfactant solution and mixed with a KPG stirrer (Janke~$\&$~Kunkel (IKA), RE16) until homogeneous. 
Subsequently, the clear liquid was poured onto a polytetrafluoroethylene tray, left to polycondensate at \qty{80}{\celsius} for 48~h, milled in a ball mill (Spex 800) and calcined at \qty{550}{\celsius} for 6~h (heating rate:~\qty{1}{\celsius\per\minute}, air flow:~\qty{4.5}{\litre\per\hour}). 
The mass ratios of the components 0.1~M HNO$_3$:TMOS:\ce{Al(NO3)3 . 9H2O}:surfactant were 0.710--0.432$\cdot$3$x$:1:3$x$:$y$, with $x$ varied from 0 for silicates to 0.164 for metallosilicates with the highest Al loading. 
No surfactant, i.e., $y = 0$, was used in the synthesis of bulk materials.
Mesoporous materials with mesopore diameters of approximately 6~nm were prepared using  Pluronic P123 (\textit{M}$_n$ = \qty{5800}{\gram\per\mole}, Sigma-Aldrich) as surfactant with $y = 0.320$. 

\subsubsection*{Transmission electron microscopy}
Samples for transmission electron microscopy were finely powdered and applied to Pioloform-coated copper grids (mesh:~200, Plano). 
Measurements were performed with an EM10C/CR TEM (Zeiss) at 60~kV. 
Images were acquired using a water-cooled 1k slow-scan CCD Camera (7888-IV, TRS) and the accompanying \textsc{ImageSP} software.

\subsubsection*{X-ray total scattering and PDF analysis}

Total scattering experiments and PDF analysis were performed as a commissioned service by Momentum Transfer GmbH. 
Measurements were performed at the ID31 beamline at the European Synchrotron Radiation Facility. 
The sample powders were loaded into cylindrical slots (approximately 1~mm thickness) held between Kapton\textsuperscript{\textregistered} windows in a high-throughput sample holder. 
Each sample was measured in transmission with an incident X-ray energy of 75.60~keV ($\lambda$ = 0.1653~\AA). 
Measured intensities were collected using a Pilatus CdTe 2M detector ($1679\times1475$ pixels, $172\times172$ $\mu$m$^2$ each) with a sample-to-detector distance of approximately 0.3~m and background corrected. 
Geometry calibration was performed using NIST SRM 660b (LaB$_6$) with \textsc{pyFAI}, followed by image integration applying flat-field, geometry, solid-angle, and polarization corrections.
The total scattering data was used to generate the Lorch-modified PDF data. 

\subsubsection*{Fourier-transform infrared spectroscopy}

Infrared spectra were recorded in transmission mode on a Nicolet 6700 spectrometer. 
Solid samples were prepared by mixing and finely grinding approximately 2~mg of the silicate material with about 50~mg of pre-dried KBr (Sigma-Aldrich, $\geq 90\%$) using a mortar and pestle, followed by pressing the mixture into pellets at a load of 2~t. Prior to pellet preparation, the silicate samples were dried under vacuum at \qty{200}{\celsius} for 24~h to remove adsorbed water. All sample handling and measurements were carried out in an argon-filled glove box to minimize moisture contamination.

\section*{Data Availability}

Data related to this manuscript are openly available in the DaRUS Repository~\cite{jung26data-permlink}. This includes the training sets (\textsc{VASP} \texttt{OUTCAR} files), the trained lightweight MLIPs, and experimental data such as Fourier-transform infrared spectra, total scattering curves, and pair distribution function data. 
Additional details on lightweight MLIP training, simulation methods, and the analysis of bulk materials and surface functional groups are provided in the Supplementary Information.

\section*{Code Availability}
Related scripts for performing simulations and analyzing results will be openly available in the DaRUS Repository (\url{https://doi.org/10.18419/DARUS-5726}) 
\cite{jung26data-permlink}.

\begin{acknowledgments}
We appreciate discussions with Yuji Ikeda, Xi Zhang, and Nikolay Zotov;  
Xiaochen Du and Mingrou Xie from MIT;
the help of Konstantin Gubaev and Julian Greif in providing parts of the training sets for silica and metallosilicates;
the help of Ruba Ajjour and Huy Bui Duc in synthesizing some of the investigated materials; 
and the assistance of Yingchun Zhang in setting up ReaxFF calculations.
We acknowledge the European Synchrotron Radiation Facility for providing measurement facilities and the Momentum Transfer GmbH for performing the total scattering experiments and providing the PDF data. 
We appreciate the fruitful discussion of the PDF analysis with Maxwell W. Terban. 

This project is supported by the Deutsche Forschungsgemeinschaft (DFG, German Research Foundation) under the Collaborative Research Centres (SFB 1333, grant No.~358283783 SFB 1333/2 2022) and under Germany's Excellence Strategy (EXC 2075-390740016). 
This project has received funding from the European Research Council (ERC) under the European Union’s Horizon 2020 research and innovation program (grant agreement No.~865855) and under the European Union’s Horizon Europe Research and Innovation Programme (Grant Agreement No.~101200433, project META-LEARN). 
The authors acknowledge support by the state of Baden-Württemberg through bwHPC, the DFG through grant No.~INST 40/575-1 FUGG (JUSTUS 2 cluster), and the Ministry of Science, Research and the Arts Baden-Württemberg.
B.G. and Y.O. acknowledge the support by the Stuttgart Center for Simulation Science (SimTech).
The authors gratefully acknowledge the scientific support and HPC resources provided by the Erlangen National High Performance Computing Center (NHR@FAU) of the Friedrich-Alexander-Universität Erlangen-Nürnberg (FAU) under the NHR project~a102cb.
NHR funding is provided by federal and Bavarian state authorities. 
NHR@FAU hardware is partially funded by the DFG grant No.~440719683. 

Funded by the European Union. 
Views and opinions expressed are, however, those of the author(s) only and do not necessarily reflect those of the European Union or the European Research Council Executive Agency. Neither the European Union nor the granting authority can be held responsible for them.

\end{acknowledgments}

\vspace{0.6cm}
\section*{Author contributions}

All authors designed the project, discussed the results, and wrote the manuscript. J.J., T.S., Y.O., B.G. performed the calculations, and S.I., J.B. performed the experiments. M.H. and N.H. discussed the results. B.G. and J.B. acquired funding.

\section*{Additional information}
\textbf{Competing interests}: The authors declare no competing interests.

\textbf{Large language model usage}: During the preparation of this manuscript, the authors used ChatGPT (OpenAI GPT-5 model) and Google Gemini (3.1 Pro Preview) to improve clarity and readability. 
All content generated with its assistance was reviewed, edited, and verified by the authors, who take full responsibility for the final content of the published article.

\onecolumngrid

\clearpage

\begin{exttable*}[h]
\centering
\caption{
\textbf{Overview of prior studies that developed \textit{ab initio}-based machine-learning interatomic potentials (MLIPs) referenced to density functional theory (DFT) data.}
We summarize studies on silicon-related systems that examined key structural motifs, including amorphous matrices, ordered mesoporosity, surface functionality, and interfacial water.
The present work extends prior studies by addressing a broader and more complex range of structural motifs. 
}
\label{table:literature}
\centering
\small
\begin{ruledtabular}
\begin{tabular}{@{}ccccccccc@{}}
 & 
 &
 &
 &
\multicolumn{4}{c}{Structural motif} & \\ 
\cmidrule(lr){5-8}
 \multirow{1}{*}{Year} & 
 \multirow{1}{*}{System} & 
 \multirow{1}{*}{XC functional$^\mathrm{a}$} & 
 \multirow{1}{*}{MLIP$^\mathrm{b}$} &
\makecell{Amorphous \\ matrix} &
\makecell{Ordered \\ mesoporosity} &
\makecell{Surface \\ functionality} &
\makecell{Interfacial \\ water} &
\multirow{1}{*}{Reference} \\
\midrule
2018  & Si &  PW91 & GAP    & x  &   &   &  & \citen{deringer2018}    \\
2018  & Si--O & PBE & GMM   & x    &     &    &  & \citen{li2018}  \\
2021  & Si--Ca--O--H  & PBE & BPNN  &   &  &  &   x & \citen{kobayashi2021}  \\
2022  & Si--B--O--F  & PBE & DeePMD  & x  &   &   &  & \citen{urata2022} \\
2022  & Si--O & SCAN & GAP  & x  &  &   & &  \citen{erhard2022} \\
2022  & Si--H & PBE & GAP  & x & &   &  & \citen{unruh2022}   \\
2024  & Si--Na--O  & PBE & DeepPot-SE  & x & &  &  & \citen{bertani24} \\
2024  & Si--O & SCAN & ACE  & x &   &    &  & \citen{erhard2024}  \\
2024  & Si--Al--O--H$^\mathrm{c}$ & \makecell{SCAN-D3(BJ), \\ $\omega$B97X-D3(BJ)} & SchNet  &   &  &   & x & \citen{erlebach2024}  \\
2024  & Si--H--O & PBE & DeePMD  & x  &   &  & x & \citen{li2024}   \\
2024  &  oxide glass$^\mathrm{d}$  & PBEsol & DeepPot-SE  & x &  &  &  & \citen{kayano24}  \\
2024  & Si--C--O  & SCAN & ACE  & x  &  &  &  & \citen{leimeroth2024}   \\
2024  & Si--C--O--H & PBE & MTP  & x  &  &  &  & \citen{falgoust2024}   \\ 
2024  & Si--O--H  & $\omega$B97X-D3 & PaiNN  &  & &  & x & \citen{Roy_Dürholt_Asche_Zipoli_Gómez-Bombarelli_2024} \\
2025 & Si--C--O--H & optB88-vdW & DeePMD & x & x &  & & \citen{liu2025} \\
2025 & Si--Al--O--H & PBE-D3 & NequIP & x & & x & & \citen{sawant2025} \\
2026 & Si--C--O--H & PBE-D3 & GAP & & & x &   & \citen{strugovshchikov26} \\
\midrule
2026  & Si--Al--O--H & \makecell[c]{r$^2$SCAN-D4, \\ vdW-DF-cx} & \makecell[c]{lightweight \\ MTP}  & x  & x & x & x & This work \\
\end{tabular}
\end{ruledtabular}
\begin{flushleft}
\vspace{-2mm}
{ \footnotesize
$^\mathrm{a}$ Exchange--correlation (XC) functionals: Perdew--Wang (PW91)~\cite{Perdew_Chevary_Vosko_Jackson_Pederson_Singh_Fiolhais_1992}, Perdew--Burke--Ernzerhof (PBE)~\cite{Perdew_Burke_Ernzerhof_1996}, revised PBE for solids (PBEsol)~\cite{Perdew_Ruzsinszky_Csonka_Vydrov_Scuseria_Constantin_Zhou_Burke_2008}, strongly constrained and appropriately normed (SCAN)~\cite{Sun_Ruzsinszky_Perdew_2015}, regularized--restored SCAN (r$^2$SCAN)~\cite{Furness_Kaplan_Ning_Perdew_Sun_2020}, dispersion corrected r$^2$SCAN-D4\cite{ehlert21}, range-separated hybrid functionals with dispersion correction $\omega$B97X-D3\cite{Lin_Li_Mao_Chai_2013} and $\omega$B97X-D3(BJ)\cite{Asim_2018}, nonlocal van der Waals functionals vdW-DF-cx~\cite{Berland_Hyldgaard_2014} and optB88-vdW~\cite{klimes2010}. Dispersion corrections: D3\cite{Grimme_Antony_Ehrlich_2010}, D3(BJ)~\cite{Grimme_Ehrlich_Goerigk_2011}, D4~\cite{Caldeweyher_Mewes_Ehlert_Grimme_2020}. \\
$^\mathrm{b}$ Machine-learning interatomic potentials (MLIPs): Gaussian approximation potential (GAP)~\cite{Bartók_Payne_Kondor_Csányi_2010}, Gaussian mixture model (GMM)~\cite{Pham_Kino_Terakura_Miyake_Dam_2016}, Behler--Parrinello neural network potential (BPNN)~\cite{Behler_Parrinello_2007}, deep potentials (DeePMD)~\cite{Wang_Zhang_Han_E_2018}, deep potentials smooth edition (DeepPot-SE)~\cite{Zhang_Han_Wang_Saidi_Car_Weinan_2018}, atomic cluster expansion (ACE)~\cite{Drautz_2019}, SchNet~\cite{Schütt_Kindermans_Sauceda_Felix_Chmiela_Tkatchenko_Müller_2017}, polarizable atom interaction neural network (PaiNN)~\cite{Schutt_Unke_Gastegger_2021}, moment tensor potential (MTP)~\cite{shapeev16}. \\
$^\mathrm{c}$ Proton-exchanged aluminosilicate zeolites. \\
$^\mathrm{d}$ Eight-component glass: \ce{SiO2}--\ce{Al2O3}--\ce{B2O3}--\ce{Na2O}--\ce{CaO}--\ce{MgO}--\ce{ZrO2}. 
}
\end{flushleft}
\end{exttable*}

\clearpage
\begin{extfigure*}[t]  
   \centering  
    \includegraphics{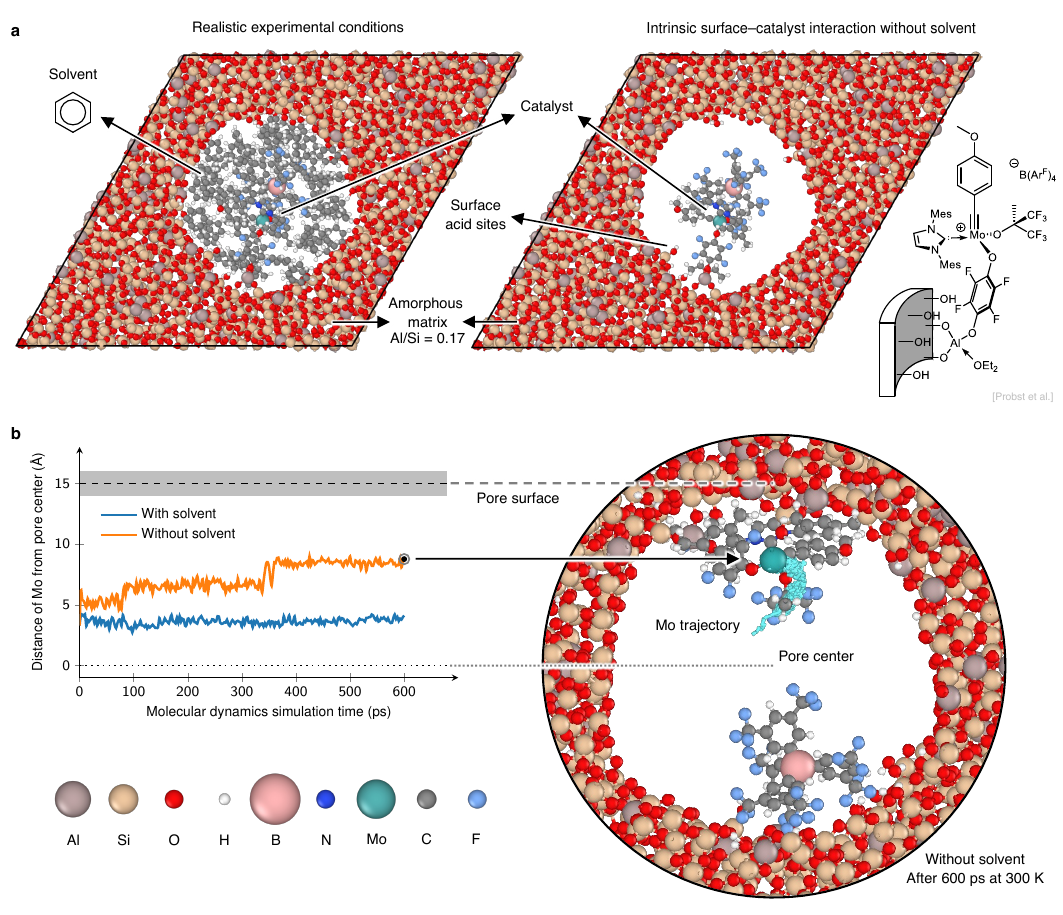}
   \caption{
   \textbf{Toward catalysis under confinement in realistic mesoporous metallosilicates.}
   \textbf{a}, Simulation setup. 
   As an example, the cationic molybdenum alkylidyne \textit{N}-heterocyclic carbene complex [Mo($\equiv$CC$_6$H$_4$-$p$-OMe)(OCMe(CF$_3$)$_2$)$_2$(IMes)][B(Ar$^{\mathrm{F}})_4$] (IMes = 1,3-dimesitylimidazol-2-ylidene, Ar$^{\mathrm{F}}$ = 3,5-bis(trifluoromethyl)phenyl) from Groos et al.~\cite{groos2021,groos2021ccdc}, used in ring-expansion metathesis polymerization of cyclic olefins to form cyclic polymers under confinement~\cite{probst2025}, is anchored within the ordered mesopores as described by Probst et al.~\cite{probst2025} 
   To investigate the intrinsic surface--catalyst interaction, the solvent can be hypothetically removed from the pore in the simulations.
   \textbf{b}, Molecular dynamics simulations.
   Simulated trajectories up to \qty{700}{ps} are provided in Video~\textcolor{blue}{S1} (with solvent) and Video~\textcolor{blue}{S2} (without solvent).
   Results are computed using the GRACE-1L-OAM foundation model for demonstration.
   In the absence of a solvent, the cationic molybdenum approaches the pore surface, suggesting a stronger influence of acid sites.
   In simulation, acid-site density can be tuned via metallosilicate composition and pore geometry, providing guidance for \textit{a priori} design.
   }
   \label{fig:catalyst}
\end{extfigure*}

\clearpage
\begin{extfigure*}[t]  
   \centering  
    \includegraphics{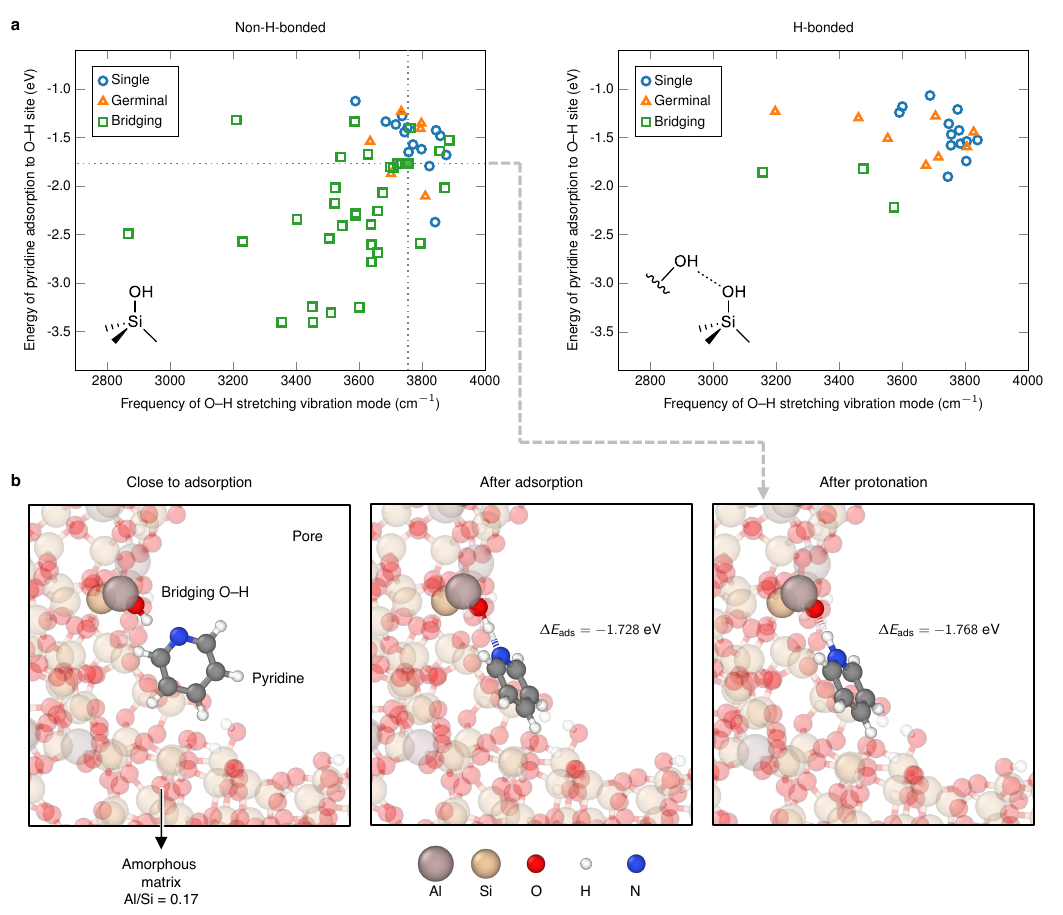}
   \caption{
   \textbf{Vibrational and adsorption properties of hydroxyl groups for the rational design of porous adsorbents.}
   \textbf{a},~Correlation between O--H stretching frequency and adsorption energy.
   Results are computed using the GRACE-2L-OAM foundation model with D3 dispersion corrections for demonstration.
   Porous metallosilicates generated by our end-to-end framework provide realistic atomic structures for evaluating O--H stretching frequencies (in the absence of adsorbates) and pyridine adsorption energies at OH sites, capturing both non-H-bonded and H-bonded OH environments~\cite{tielens20, Gierada_Petit_Handzlik_Tielens_2016}.
   \textbf{b},~Representative pyridine adsorption at \SI{0}{\kelvin} on a bridging OH site which exhibits a stretching frequency of \qty{3756}{cm$^{-1}$}.
   The full adsorption trajectory is provided in Video~\textcolor{blue}{S3}.
   Protonation upon adsorption lowers the adsorption energy, $\Delta E_{\mathrm{ads}}$, by about \qty{40}{meV}.
   These insights provide a basis for designing the surface structure of porous metallosilicates as adsorbents. 
   Computational details are provided in Supplementary Sec.~S3 C.
   }
   \label{fig:acidity}
\end{extfigure*}

\clearpage
\twocolumngrid
\bibliography{references_arxiv}

\clearpage

\ifarXiv
    \includepdf[pages={{},-}, link=true]{\supplementfilename}
\fi


\end{document}